\documentclass[aps,showpacs,floatfix]{revtex4}

\usepackage{amsmath}
\usepackage{graphicx}

\begin{document}


\title{Universal properties of high-frequency spectra in glass
  formers} 
\author{P. Verrocchio} 
\affiliation{Dipartimento di Fisica, Universita' di Trento Povo (TN)
  Italia}
\begin{abstract}
  The vibrational spectra of glass formers follow different laws with
  respect to crystals. A rationale for their anomalous behaviour is
  provided by the euclidean random matrix theory.  Experiments on
  glass formers at different densities might be a suitable route to
  test the prediction of the theory.
\end{abstract}
\date{\today}
\pacs{61.43.-j 61.43.Fs}
\maketitle


\section{Introduction}
The high frequency ($Thz$) dynamics of the density fluctuations in
highly viscous supercooled systems is presumably due to purely
vibrational modes. However, with respect to their crystalline
counterpart these systems show a number of puzzling feautures whose
theoretical interpration is still controversial (see
e.g.~\cite{pisa02}). The purpose of this work is to focus on two of
them describing a theoretical approach which is able both to explain the
physical processes beyond such features and to make rigourous
predictions.
\begin{description}
\item[Brillouin-Peak] $X$-rays and Neutron inelastic scattering
 experiments probe excitations of momentum $q$ comparable to the
 characteristic momentum $q_0$, the position of the first diffraction
 peak in the static structure factor. In other words the wavelength is
 of the same order of magnitude of the inter-particle distance and an
 hydrodynamic approximation does not make much sense. Nevertheless the
 position $\omega$ and the width $\Gamma$ of the Brillouin Peak are
 reported to show {\em hydrodynamic}-like features:
 \begin{equation}
   \omega_{peak} \propto q,  \hspace{.5cm} \Gamma_{peak} \propto q^2 
   \label{BRILLOUIN}
 \end{equation}
 The agreement regarding the exponent $2$ in the scaling law of
 $\Gamma$ between hydrodynamics and experiments is likely
 fortuitous. In fact the magnitude of $\Gamma$ has a very slight
 dependence on temperature (if any) while it should be strongly
 temperature dependent in case the decay of hydrodynamic modes was
 induced by the viscosity~\cite{Hansen86}.
\item[Boson Peak] Raman experiments and neutron scattering experiments
  at large $q$ (see below) may be useful to determine the vibrational
  density of states (VDOS) $g(\omega)$. Glass formers are quite
  peculiar with respect to crystals since in the frequency region
  where the main hypothesis of Debye holds (linearity of the
  dispersion relation) one finds that the VDOS has a different
  behaviour from the Debye law:
  \begin{equation}
    g(\omega)/g_{Debye}(\omega) >> 1.
    \label{BOSON}
  \end{equation}
  This excess of states over the Debye VDOS is termed {\em Boson
  Peak} (BP).
\end{description}

\section{Theory}

Within the harmonic framework one assumes that the VDOS is simply
given by the eigenvalues of the Hessian Matrix ${\cal K}_{i j ; \mu
\nu}$.  At low enough frequency this approximation breaks down since
the relaxational modes start playing a relevant role. One could
roughly claim that the harmonic approach holds whenever $\omega$ is
larger than the inverse of the {\em fastest} relaxation mode
time-scale. Since this quantity in all the realistic cases is unkonwn,
one must content to assess empirically the range of reliability of the
harmonic approach by means of the comparison between the experimental
spectra and the numeric spectra~\cite{Pilla04}. Aiming to modelize the
vibrations in glass-formers one has to take into account the role
played by the disorder. As a matter of fact the positions around wich
the atoms vibrate harmonically do not form in general a regular
lattice. The disorder of the positions $\{\vec x_i\}$ of the center of
oscillation is described by a suitable distribution of probability
$P[\{\vec x_i\}]$. Different choices for $P$ describe different
physical situations. Furthermore this topological disorder induces a
broad distribution of probability of the spring constant ${\cal
K}$. The theoretical effort is greatly simplified if the macroscopic
quantities are assumed to be {\em self-averaging}, i.e if their value
remains the same when different realizations of the disorder are
taken~\cite{spinglass98}. This allows to replace the computation of
(say) the VDOS for a given realization of the disorder (involving the
knowledge of an infinite number of positions $\vec x_i$) with the
computation of the averaged VDOS (which involves the knowledge of
$P$). Besides the VDOS another interesting quantity is the Vibrational
Dynamic Structure Factor (VDSF) $S(q,\omega)$, which shows the
Brillouin Peak.  The two quantities are written as:
\begin{equation}
  \displaystyle 
  g(\omega) \equiv \frac{1}{N} \overline{\sum_n
  \delta(\omega-\omega_n)} \quad 
  S(q,\omega) \equiv \frac{k T}{N \omega^2} \overline{\sum_n
  \left|\sum_i \vec q \cdot \vec e_{n,i} \mathrm{e}^{\mathrm{i} \vec q
  \cdot \vec x_i } \right|^2 \delta(\omega-\omega_n)} 
\end{equation}
where $\vec e_n$ are the eigenvectors and $\omega^2_n$ the eigenvalues
of the Hessian matrix ${\cal K}$. The above expression of the VDSF
arises when only $1$-phonon process are considered. In this work we
address the computation of such quantities for {\em topologically
disordered models}, which we claim to be the most appropriate to
describe glass formers.

In order to compute the spectral properties of the matrix ${\cal K}$
\begin{equation}
K_{ij;\mu \nu} \equiv \delta_{ij}\sum_k \partial_{\mu
\nu} V(|\vec r_{ik}|) - \partial_{\mu \nu} V(|\vec r_{ij}|)
\hspace{1cm} \vec r_{ij} \equiv \vec x_i- \vec x_j.
\label{KAPPA}
\end{equation}
we introduce a complex valued tensor $G_{\mu \nu}$ called {\em resolvent},
which can be split in its longitudinal and transversal parts:
\begin{eqnarray}
\displaystyle G_{\mu \nu}(z,q) \equiv \frac{1}{N} \overline{\sum_{jk}
\mathrm{e}^{\mathrm{i} \vec q \cdot \vec r_{jk}} \left[
\frac{1}{z-K}\right]}_{jk; \mu \nu} \equiv G_L(z,q) \frac{q_\mu
q_\nu}{q^2} + G_T(z,q) \left( \delta_{\mu \nu} - \frac{q_\mu
q_\nu}{q^2} \right)
\end{eqnarray}
$z$ is a complex number related to the value $\lambda$ of the
eigenvalues of the Hessian matrix ${\cal K}$ by $z \equiv \lambda +
\mathrm{i} \epsilon$.

The resolvent is related to the VDOS and the VDSF. However $G_{\mu
\nu}$ is defined in the space of eigenvalues which nothing prevents
from being negative. On the other hand the VDSF and the VDOS are
functions of the frequency, and the frequencies are related to the
eigenvalues only in the positive eigenvalues region ($\lambda =
\omega^2$). Then the following relations hold only in that region:
\begin{eqnarray}
\displaystyle 
S(q,\omega) = - \frac{2 k T q^2}{\omega \pi} {\rm Im} \,
G_L(q,\omega^2+ \mathrm{i} 0^+), \hspace{.5cm}
g(\omega)=\lim_{q \to \infty} \frac{\omega^2}{k T q^2} S(q,\omega) 
\label{FREQUENZE}
\end{eqnarray}
However we will show that the existence of the resolvent at negative
$\lambda$ leads to important conseguences. Note that only {\em
isotropic} modes survives in the VDOS because $G_{\mu \nu}(q=\infty,z)
\propto \delta_{\mu \nu}$. Thus the polarization is approximately
defined only for $q \sim 0$. Ihis is very different from lattice
models, where $g(\omega) \propto \int d \vec q S(q,\omega)$.

The resolvent can be computed in a non perturbative way if the $n$-th
moment of $P$ is assumed to be factorized ({\em superposition
approximation}): $g^{(n)}(\vec x_1 \dots \vec x_n) = g(\vec r_{12})
\dots g(\vec r_{n-1 \, n})$ where $g(r)$ is the pair probability
distribution. With this approximation the resolvent is given
by~\cite{Ciliberti05}:
\begin{equation}
  \displaystyle 
  G_{\mu \nu}(q,z) = \left[\frac{1}{z-\rho\hat{f}(0)+\rho\hat{f}(q)-
  \Sigma(q,z)}\right]_{\mu \nu} \quad
  \Sigma(q,z)_{\mu \nu} =
  \frac{1}{\rho} \sum_{\alpha \beta} \int \frac{d^3 k}{(2 \pi)^3}
  V_{\mu \alpha}(q,k) G_{\alpha \beta}(k,z) V_{\beta \nu}(k,q)
  \label{RESOLVENT}
\end{equation}
where $\rho$ is the density, $V_{\mu \nu}(k,q) \equiv
\rho\left(\hat{f}_{\mu \nu}(k))-\hat{f}_{\mu \nu}(q-k)\right)$ and
$\hat{f}_{\mu \nu}(q)\equiv {\cal F}[g(r) \partial_{\mu \nu} v(r)$
($v(r)$ is the pair interaction).

The quantity $\Sigma$ is called {\em self-energy} and describes the
loss of energy of phonons due to the disorder. It vanishes in the
limit of infinite density, where phonons propagate without
dissipation.  In this limit one has:
\begin{equation}
  S(q,\omega) = \delta \left( \omega - \omega^0_{peak}(q) \right)
 \quad \omega^0_{peak}(q) \equiv \sqrt{\rho\left(\hat f(0) - \hat
 f_L(q) \right)}
  \label{FREE}
\end{equation}
This does not imply that the disorder vanishes when $\rho \to \infty$.
Rather his effects on the propagations of phonons are {\em averaged
out} by the infinite number of particles within each wavelength. On
the other hand at finite $\rho$ the self-energy starts playing a
relevant role, modifying the position and the width of the peak.

\section{Evolution of the Boson Peak with density}

Solving the integral equation~(\ref{RESOLVENT}) in the limit of $q \to
\infty$ one obtains the density of eigenvalues from which the VDOS
$g(\omega)$ is obtained. Due to the complexity of the equation, the
result for a given choice of $\hat{f}_{\mu \nu}(q)$ can be computed
only numerically. However it is possible to deduce some important
mathematical properties of the solution which are not dependent on
$\hat f$ (hence, on the model). Keeping fixed the other thermodynamic
parameter, it can be shown that at low enough densities the Hessian
matrix has both negative and positive eigenvalues, while at high
densities all the eigenvalues are positive. A critical density
$\rho_c$ separates the two regions. In the low density
regime~(\ref{FREQUENZE}) implies that $g(\omega) \propto \omega$ when
$\omega \sim 0$ while at very high densities one has the standard
Debye behaviour $g(\omega) \propto \omega^2$. The change of exponent
in $g(\omega)$ is due to the occurence of a sort of phase transition.
The parameter which induces such phase transition is not necessarily
the density, a changement in any of the other thermodynamic quantities
may lead the system from the region with only positive eigenvalues of
${\cal K}$ to the region where also the negative ones are allowed. For
example it has been shown~\cite{Grigera03} that the value of the
potential energy of the stationary points (minima or saddles) around
which a glass system {\em remains} for a long time (termed generalized
inherent structures) are one of those quantities. For the experiments
however the stationary points are not a suitable observable, while it
is surely conceivable perform measurements of spectra at different
densities putting the system under a very high pressure. In the
following we will identify the region without negative eigenvalues
with the glass forming liquid. Let us see in detail the behaviour of
the VDOS in this phase. In the region of frequencies where the
dispersion relation is still linear one finds the following universal
form:
\begin{eqnarray}
  g(\omega)=\omega^\gamma h(\omega\Delta^{-\alpha}),
  \qquad h(x)\sim \left\{
  \begin{array}{lcr}
    x^{2-\gamma}&&x\ll 1\\
    \textrm{const.}&&x\gg 1
  \end{array}
  \right.
  \label{SOLUZIONE}
\end{eqnarray}
If we define a {\em distance} $\Delta$ from the critical point as
$\Delta \equiv (\rho-\rho_c)$ we see that at a charateristic frequency
$\omega_{BP}$ it occurs a crossover from the Debye behviour $g(\omega)
\sim \omega^2$ to a different law $g(\omega) \sim \omega^\gamma$. One
obtains furthermore that:
\begin{equation}
  \omega_{BP}\sim \Delta^\alpha, \qquad
  g(\omega_{BP})/g_{Debye}(\omega_{BP}) \sim\Delta^{-\eta}
  \label{BP}
\end{equation}
where the universal exponents are:
\begin{eqnarray}
  \begin{array}{|c|c|c|}
    \eta & \alpha & \gamma \\ \hline 1/2 & 1 & 3/2
  \end{array} 
  \label{ESPONENTI}
\end{eqnarray}
\begin{figure}
  \includegraphics[height=7.5cm,angle=270]{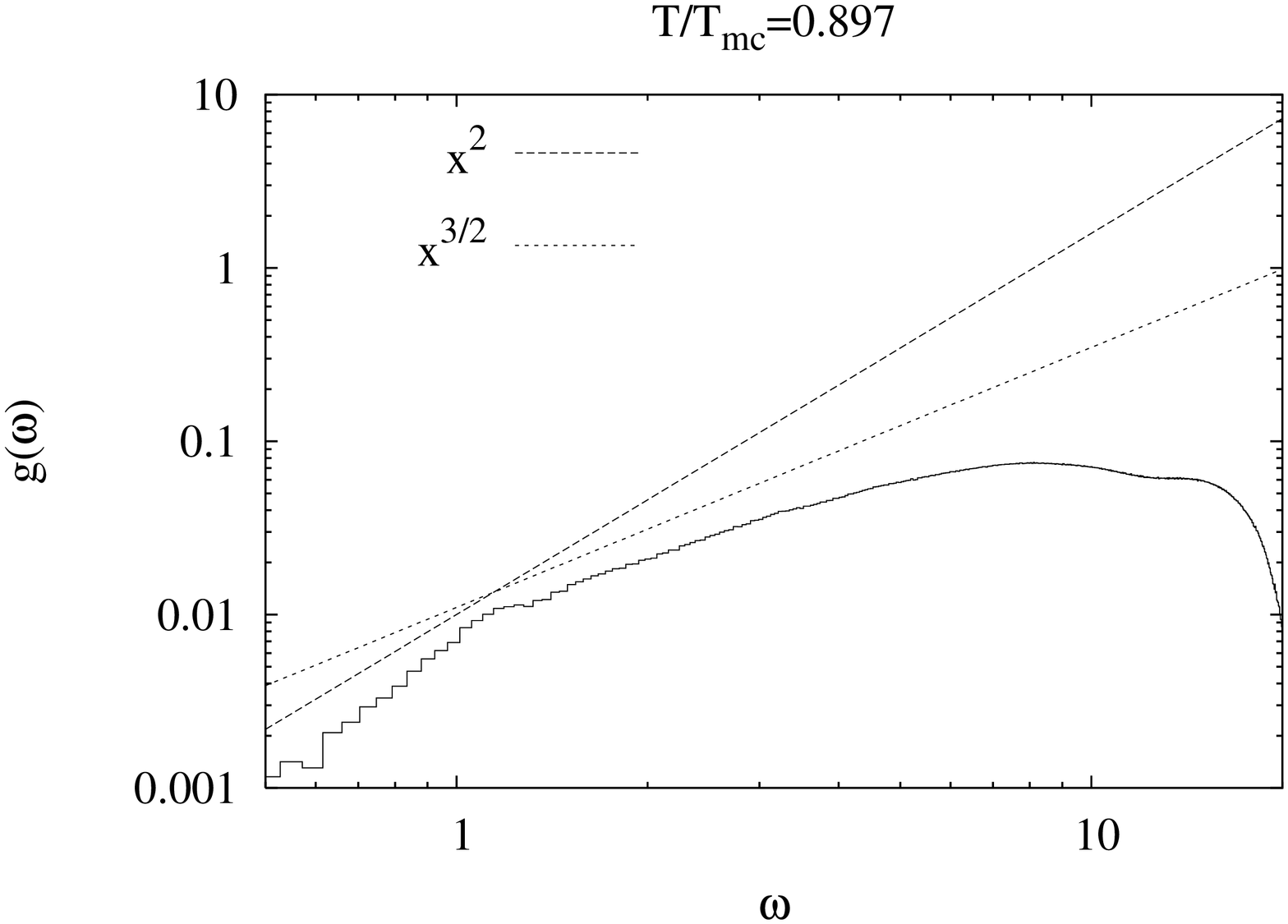}
  \includegraphics[height=7.5cm,angle=270]{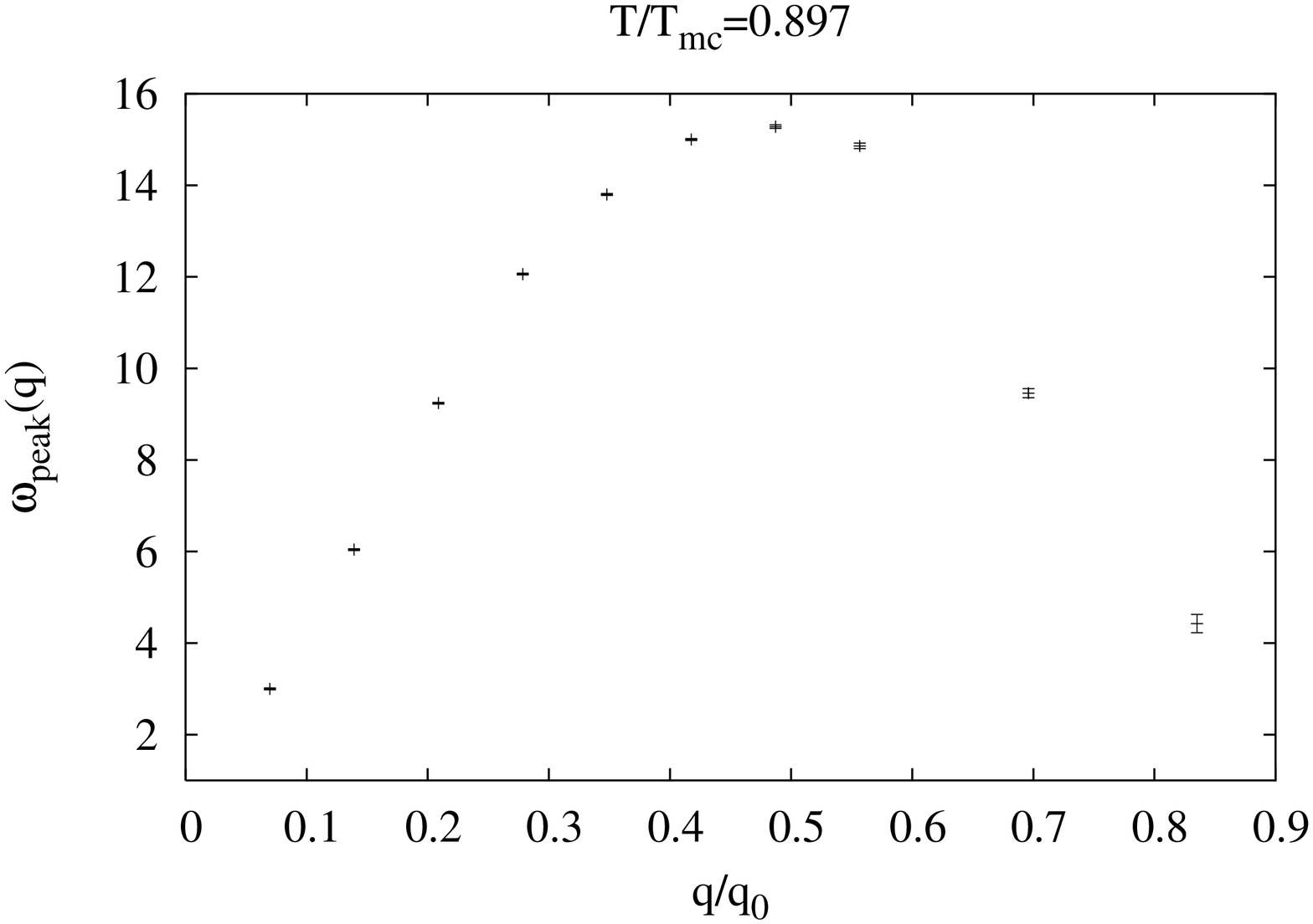}
  \caption{Left: The VDOS in log-log representation compared with
    the $\omega^2$ and $\omega^{3/2}$ laws predicted by the
    theory. Right: The position of the Brillouin Peak as a function
    of the momentum $q$.}
  \label{CAMBIO}
\end{figure}
The main conseguence of the a critical point is the modification of
the Debye law at $\omega_{BP}$. This is higly remindful of the Boson
Peak found in experiments. This suggests to identify $\omega_{BP}$
with the position of the Boson Peak and
$g(\omega_{BP})/g_{Debye}(\omega_{BP})$ with its
height. Since~(\ref{SOLUZIONE}) does not hold at low $\omega$ the
predicted singularity at $\rho_c$ is not approached in real
systems. However the proximity of such singularity is enough to break
the Debye law.

We can test our theoretical prediction by performing numeric
simulations of a simple glass former. We performed MonteCarlo
simulations at different temperatures and densities of a simple glass
former (see~\cite{Grigera04} of the details). The VDOS's have been
computed by diagonalizing at least $50$ different realizations of the
Hessian matrix.

\begin{figure}
  \includegraphics[height=7.5cm,angle=270]{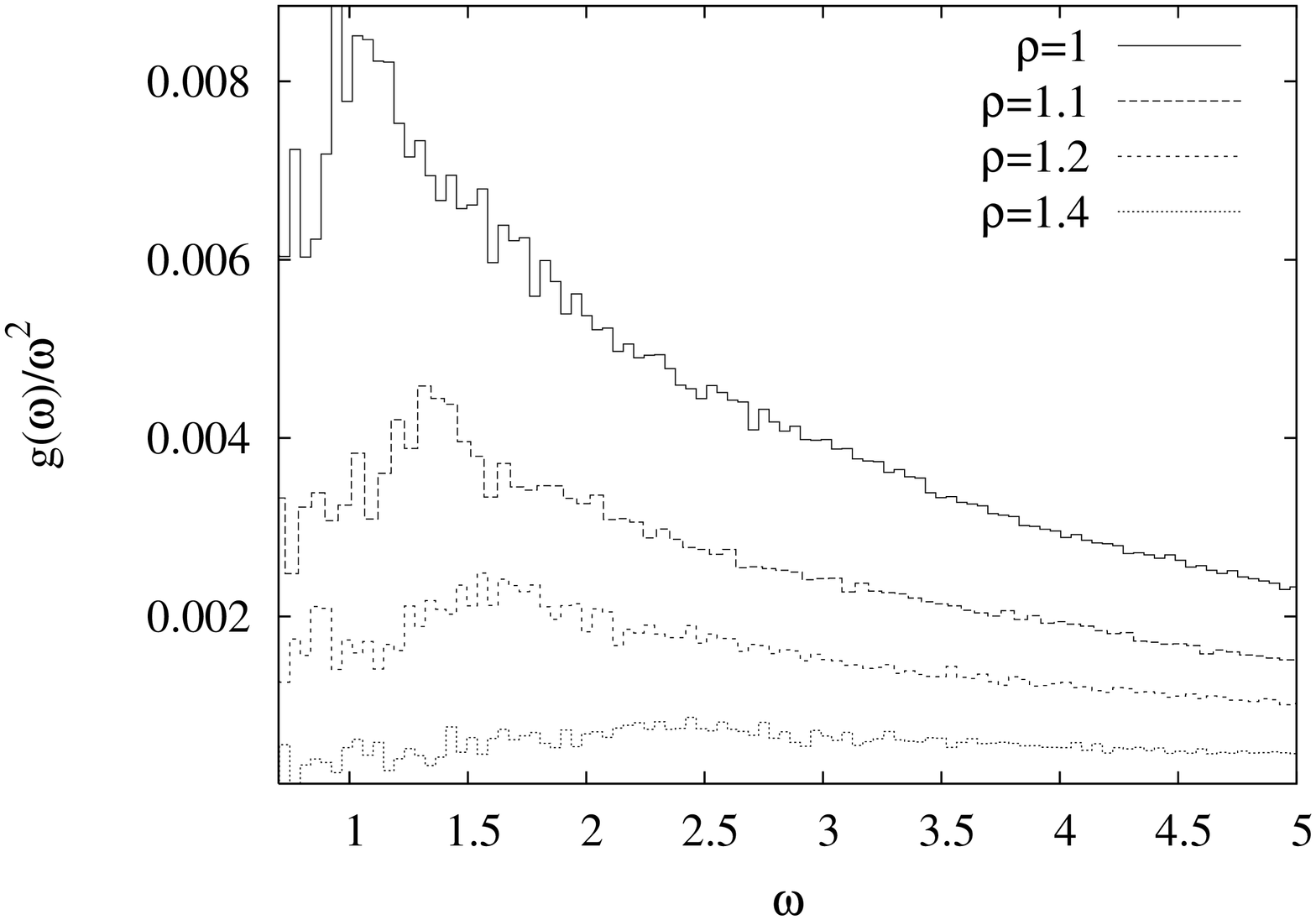}
  \includegraphics[height=7.5cm,angle=270]{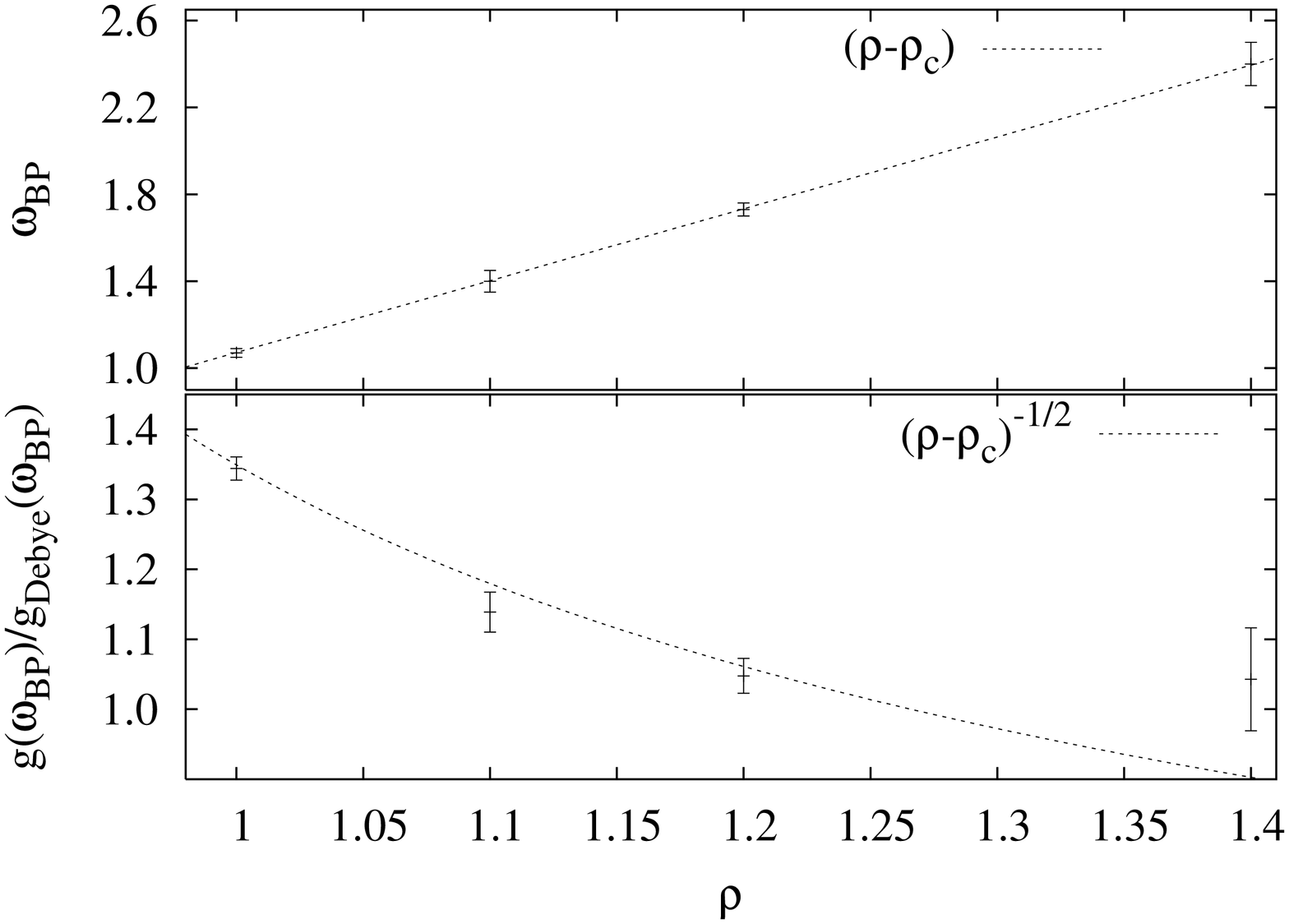}
  \caption{Left: VDOS divided by $\omega^2$ at different
    $\rho$. Right: The position and the heigth of the BP as a
    function of the density}
  \label{DENSITA}
\end{figure}

In fig~(\ref{CAMBIO}) we show the VDOS for $T = 0.897 T_\mathrm{mc}$
($T_\mathrm{mc}$ is the Mode Coupling temperature~\cite{goetze92}).
At $\omega_{BP} \sim 1$ (where the dispersion relation is still
linear) there is the crossover in the VDOS from the Debye law to the
$\omega^{3/2}$ behaviour. This is the signature of the Boson Peak, as
predicted by the theory.  In fig~(\ref{DENSITA}) we see the large
variations of the VDOS for different densities at $T = 4.420
T_\mathrm{mc}$ (left). The position of the Boson Peak moves linearly
with the density while its the growth of its eight seems to be
compatible with a power law whose exponent is $1/2$ (right).  Thus
even those theoretical predictions are in reasonable agreement with
the numeric data.

\section{The complex behaviour of the Brillouin peak}
At finite densities the propagation of phonons is affected by the loss
of energy due to the disorder which make the time-life of phonons
finite. Actually it is not entirely correct to describe the effect of
disorder on the phonons in topologically disordered systems in terms
of phonon-disorder interaction. In fact as we said above in the
infinite density limit the phonons propagate freely even if the
positions of the particles remain disordered. Actually, when the
disorder has a topological nature it is not possible to find a limit
where the disorder vanishes, even if his effects may disappear. This
is at odds with lattice systems, where phonons propagate freely only
when the disorder is not present. The position and the width of the
Brillouin peak are affected by the self-energy in the following way:
\begin{eqnarray}
  \omega_{peak}(q)& \sim & \sqrt{\rho\left(\hat f(0) - \hat
    f_L(q)\right) + {\rm Re} \Sigma_L(\omega^0_{peak})} \\
    \Gamma_{peak}(q)& \sim & \frac{{\rm Im}
    \Sigma_L(\omega^0_{peak})}{\omega}
  \label{PICCO}
\end{eqnarray}
Aiming to obtain model independent results we limit to the asymptotic
region where $q$ is small with respecto to $q_0$. The general result
is:
\begin{eqnarray}
  \displaystyle \omega_{peak}(q) &\sim& {c \, q}
  \label{POSIZIONE} \\
  \Gamma_{peak}(q) &\sim& A \,
  {\frac{g(\omega_{peak})}{{\omega_{peak}}^2} \, q^2 + B \, q^4}
  \label{ALLARGAMENTO}
\end{eqnarray}
($c$ is the speed of sound and $A,B$ two model dependent
constants). While the eq.~(\ref{POSIZIONE}) signals trivially the
existence of propagating phonons, eq.~(\ref{ALLARGAMENTO}) reveals a
quite rich behaviour of the width of the Brillouin peak. When the
frequency of the Brillouin Peak lie below $\omega_{BP}$ the leading
term of $\Gamma_{peak}$ is $\propto q^2$ while at larger frequencies
there is a crossover to $\Gamma_{peak} \propto q^{3/2}$. Then
$\omega_{BP}$ signals not only the frequency with the maximum excess
of states with respect to the Debye law but even a crossover in the
power law of the width of the Brillouin Peak. Moreover at large enough
momenta there exists another crossover frequency $\omega_\star$ where
the $q^4$ term becomes dominant.

We have seen in the previous section that for very large densities the
BP moves a very large frequencies hence it is reasonable to expect
that in this situation one should observe only the crossover at
$\omega_\star$ from the $\Gamma \propto q^2$ to $\Gamma \propto
q^4$. 
Not only $\omega_{BP}$ changes with the density (see eq.~(\ref{BP})),
even the characteristic frequency $\omega_\star$ does, though with a
different behaviour. At high densities $A \propto 1/\rho^2, \, B
\propto 1/\rho$ hence $\omega_\star \propto 1/\sqrt{\rho}$ at $\rho >>
1$~\cite{Ciliberti05}.

\section{Conclusions}
The euclidean random matrix theory provides a coherent description of
the anomalous features found in the vibrational spectra of glass
formers. It provides furthermore well defined quantitative laws which
should be compared with the numeric and experimental findings.  In
this paper we presented mainly the laws describing the evolution of
the Boson Peak and of the Brillouin Peak when the density of the
system is changed.

How many chances do the experiments have to investigate such laws?
The main restraint we see is given by the limited frequency windows
where the harmonic modes describe entirely the dynamics. In fact
already in the $Ghz$ region the relaxational modes (accounted for
example by the Mode Coupling theory) could become
important. Optimistically we might argue that the harmonic approach
holds at most over three decades (say for frequencies $\in [0.01-10]
Thz$). This raises some problems in the verification both of
eq.~(\ref{BP}) and of eq.~(\ref{ALLARGAMENTO}). In the former case one
cannot get very close to the critical point underlying the Boson Peak
since the low frequency region is not described by harmonic
effects. Still, the numeric results in fig.~(\ref{DENSITA}) suggest
that the singularity might be discerned and studied quantitatively
even in experiments.  In the latter case the possible presence of two
crossovers is hard to detect having only three decades of frequencies
available. Not surprisingly different experimental groups interpret
the results in this region in different
ways~\cite{Benassi96,Foret97}. In that kind of meausrement it might be
useful to remind that playing with density is it possible to change
the values of $\omega_{BP}$ and $\omega_\star$ exalting a particular
behaviour with respect to the others.

\section{acknowledgments}
  I would like to thank S. Ciliberti, T.Grigera, M. Mezard,
  V. Martin-Mayor and G.Parisi for their help in these years. During my
  stay in Madrid my work was supported by the European Commission
  (contract MCFI-2002-01262) and by MEC (Spain), through contracts
  BFM2003-08532, FIS2004-05073 and FPA2004-02602.

\bibliographystyle{plain}
\bibliography{biblio}

\end{document}